\documentclass[final,5p,times,twocolumn,sort&compress]{elsarticle}

\usepackage{amsmath,amssymb,graphicx,MnSymbol}
\usepackage{booktabs}
\usepackage{multirow}
\usepackage{rotating}
\usepackage{sidecap}
\usepackage{floatrow}
\newcommand{\sss}{\scriptscriptstyle}
\newcommand{\sst}{\scriptstyle}

\newcommand{\stext}[1]{\sss \text{#1} \sst}

\journal{Thin Solid Films}

\begin{document}

\begin{frontmatter}



\title{Characterization of highly anisotropic three-dimensionally nanostructured surfaces}


\author{Daniel Schmidt}
\address{Department of Electrical Engineering and Center for Nanohybrid Functional Materials, University of Nebraska-Lincoln, Lincoln, Nebraska 68588-0511, USA}

\begin{abstract}

Generalized ellipsometry, a non-destructive optical characterization technique, is employed to determine geometrical structure parameters and anisotropic dielectric properties of highly spatially coherent three-dimensionally nanostructured thin films grown by glancing angle deposition. The (piecewise) homogeneous biaxial layer model approach is discussed, which can be universally applied to model the optical response of sculptured thin films with different geometries and from diverse materials, and structural parameters as well as effective optical properties of the nanostructured thin films are obtained.
Alternative model approaches for slanted columnar thin films, anisotropic effective medium approximations based on the Bruggeman formalism, are presented, which deliver results comparable to the homogeneous biaxial layer approach and in addition provide film constituent volume fraction parameters as well as depolarization or shape factors. Advantages of these ellipsometry models are discussed on the example of metal slanted columnar thin films, which have been conformally coated with a thin passivating oxide layer by atomic layer deposition. Furthermore, the application of an effective medium approximation approach to \emph{in-situ} growth monitoring of this anisotropic thin film functionalization process is presented. It was found that structural parameters determined with the presented optical model equivalents for slanted columnar thin films agree very well with scanning electron microscope image estimates.

\end{abstract}

\begin{keyword}
generalized ellipsometry  \sep glancing angle deposition \sep atomic layer deposition \sep sculptured thin films \sep biaxial anisotropy \sep effective medium approximations

\end{keyword}

\end{frontmatter}


\section{Introduction}
With sophisticated deposition techniques and growth processes it is possible to bottom-up fabricate self-organized three-dimensional nanostructures, which render an artificial material class with intriguing optical, magnetic, mechanical, electrical or chemical properties. One of these technologies is a physical vapor deposition process called glancing angle deposition, which, due to the particular growth geometry and conditions combined with dynamic substrate movement, allows for \emph{in-situ} sculpturing of self-organized, highly spatially coherent, three-dimensional achiral and chiral geometries at the nanoscale. The resulting sculptured thin films (STFs) exhibit columnar characteristics and physical film properties can be tailored by choice of material and controlling nanostructure geometry and film porosity~\cite{Hodgkinsonbook,LakhtakiaBook,Hawkeye2007}. In subsequent fabrication steps the nanostructure scaffolds may be further enhanced by surface functionalization. Atomic layer deposition (ALD) is an excellent technique to conformally coat such complex nanostructures with protective oxide coatings and ferromagnetic shells, for example~\cite{Schmidt2012a,Albrecht2010}.

Such engineered nanostructured materials constitute a new realm of solid state materials, and carry a huge potential for applications in the fields of nano-photonics~\cite{Toader2001}, nano-electromechanics~\cite{Singh2004}, nano-electromagnetics~\cite{Bell2005}, nano-magnetics~\cite{Schmidt2010a,Schmidt2013}, nano-sensors~\cite{Kesapragada2006,Steele2008,Rodenhausen2012,Hofmann2012}, and nano-hybrid functional materials~\cite{Kasputis2013}.

In order to systematically utilize STFs in future applications, physical properties of these nanosized objects need to be evaluated and understood such that targeted geometry engineering with tailored properties from selected materials and material combinations will be possible. Non-invasive and non-destructive optical techniques are preferred, however, due to the complexity of STFs, optical characterization is a challenge. Spectroscopic generalized ellipsometry within the Mueller matrix formalism is a polarization-dependent linear-optical spectroscopy approach and provides an excellent tool to determine the dielectric functions of anisotropic optical systems. Generalized ellipsometry has been shown to be an excellent optical technique to determine anisotropic optical properties of STFs of arbitrary geometry and materials upon analyzing the anisotropic polarizability response~\cite{Schmidt2013ellinanoscale}. Structural parameters such as thickness and void fraction can be derived from best-match model analysis~\cite{Kaminska2005,Beydaghyan2005,Nerbo2010}. It is also possible to determine multiple film constituents within slanted columnar thin films (F1-STFs) and this has been recently shown for thin conformal passivation layers grown by ALD and \emph{in-situ} quantification of organic adsorbate attachment analysis~\cite{Schmidt2012a,Rodenhausen2012}.

However, since ellipsometry is an indirect measurement technique, adequate optical models have to be chosen to evaluate experimental data in order to obtain reliable optical and structural properties of anisotropic samples. The film structure of metal STFs, which are in the simplest case homogeneous anisotropic lossy composite materials consisting of slanted columns of regular shape and common orientation (F1-STFs), induces form-birefringence and dichroism. Appropriate mixing formulas and effective medium homogenization approaches need to be applied to calculate an effective anisotropic dielectric medium response that renders the effects of the measured anisotropy~\cite{Schmidt2013ellinanoscale,Shivolabook}.

In case of a biaxially anisotropic composite material, the classic ellipsometry model approach is to individually determine the three major axes dielectric functions without any implications on the kind of constituents and constituent fractions of the composite. This homogeneous biaxial layer approach can deliver structural information from a thickness parameter and Euler angles~\cite{Schmidt2009a,Schmidt2009b}. If constituent fractions and information about the shape of the constituents are desired results, homogenization approaches based on Bruggeman, for example, can be applied such that the three major axes dielectric functions can be constructed from a composite model that describes the effects of shape, average constituent fraction and the use of constituent bulk-like optical constants for the materials of the buildings blocks (in general ellipsoidal inclusions)~\cite{Schmidt2013b}.

The objective of this manuscript is to briefly summarize optical model strategies to analyze the polarization-sensitive optical response of ultrathin STFs with simple and complex geometries based on the homogeneous layer approach. The example of cobalt F1-STFs conformally coated with alumina by ALD is used as a reference to illustrate how two different generalized effective medium approximations derived from Bruggeman's original formalism compare with the homogeneous layer approach and estimates obtained from scanning electron microscopy (SEM) images. Furthermore, \emph{in-situ} growth monitoring of conformal oxide coatings on permalloy (Ni$_{80}$Fe$_{20}$) F1-STFs by analysis of Mueller matrix spectra is presented.


\section{Generalized Ellipsometry}
Generalized ellipsometry (GE), a non-destructive and non-invasive optical technique, has proven to be highly suitable for determining optical and structural properties of highly anisotropic nanostructured films from metals such as F1-STFs or even helical (chiral) STFs~\cite{Schmidt2009a,Schmidt2009c,Schmidt2013ellinanoscale}. Measurements of the complex ratio $\rho$ of the $s$- and $p$-polarized reflection coefficients are presented here in terms of the Stokes descriptive system, where real-valued Mueller matrix elements $M_{ij}$ connect the Stokes parameters before and after sample interaction~\cite{HOE,Schubert2006a}.
The linear polarizability response of a nanostructured thin film due to an electric field $\mathbf{E}$ is a superposition of contributions along certain directions: ${\bf P}= \varrho_{a}(\bf{a}\cdot\bf{E})\bf{a} + \varrho_{b}(\bf{b}\cdot\bf{E})\bf{b} +\varrho_{c}(\bf{c}\cdot\bf{E})\bf{c}$~\cite{Dressel2008}.
In the laboratory Cartesian coordinate system the F1-STF is described by the second rank polarizability tensor $\chi$ and $\mathbf{P} = (\boldsymbol{\varepsilon} - 1) \mathbf{E}=\boldsymbol{\chi} \mathbf{E}$. The Cartesian coordinate system $(x,y,z)$ is defined by the plane of incidence $(x,z)$ and the sample surface $(x,y)$. This Cartesian frame is rotated by the Euler angles ($\varphi, \theta, \psi$) to an auxiliary system $(\xi, \eta, \zeta)$ with $\zeta$ being parallel to $\bf{c}$~\cite{HOE,Schubertbook}. For orthorhombic, tetragonal, hexagonal, and trigonal systems a set of $\varphi, \theta, \psi$ exists with $\boldsymbol{\chi}$ being diagonal in $(\xi, \eta, \zeta)$. For monoclinic and triclinic systems an additional projection operation ${\bf U}$ onto the orthogonal auxiliary system $(\xi, \eta, \zeta)$ is necessary, which transforms the virtual orthogonal basis into a non-Cartesian system~\cite{Graefbook}:
\begin{equation}
\label{eqi5}\mathbf{U}=\left( \begin{array}{ccc}
\sin{\alpha} &  \frac{\cos{\gamma}-\cos\beta \cos\alpha}{\sin\alpha} & 0\\
0 & [\sin^2{\beta}- (\frac{\cos{\gamma} - \cos\beta\cos\alpha}{\sin{\alpha}})^2]^{\frac{1}{2}} & 0\\
\cos\alpha & \cos\beta & 1\\
\end{array} \right).
\end{equation}
Additional internal angles $\alpha, \beta, \gamma$ are introduced into the analysis procedure, and which differentiate between orthorhombic ($\alpha=\beta= \gamma=90^{\circ}$), monoclinic ($\beta \ne 90^{\circ}$), or triclinic ($\alpha \ne \beta \ne \gamma$) biaxial optical properties.

Ellipsometric data analysis for anisotropic thin film samples requires nonlinear regression methods, where measured and calculated GE data are matched as close as possible by varying appropriate physical model parameters~\cite{HOE,Schubertbook}. The quality of the match between model and experimental data can be measured by the mean square error (MSE)~\cite{Johs1993}.
The major axes polarization response functions $\varrho_a, \varrho_b, \varrho_c$ can be extracted on a wavelength-by-wavelength basis, i.e., without physical lineshape implementations and Kramers-Kronig consistency tests can then be done individually for dielectric functions along each axis~\cite{Dressel2008}. However, a generally more robust procedure is matching parameterized model dielectric functions to experimental data simultaneously for all spectral data points. Parametric models further prevent wavelength-by-wavelength measurement noise from becoming part of the extracted dielectric functions and greatly reduce the number of free parameters.

\section{Homogeneous Biaxial Layer Approach}
The homogeneous biaxial layer approach (HBLA) assumes that a given composite material can be described as a homogeneous medium whose anisotropic optical properties are rendered by a spatially constant dielectric function tensor. This dielectric function tensor must be symmetric since no magnetic or other non-reciprocal properties are considered. The dielectric function tensor, in general, comprises three effective major axes dielectric functions $\varepsilon_j=1+\varrho(\omega)_{j}$ as described in Eq.~(\ref{eq:monoclinicepstensor}), and may represent an anisotropic material resembling either orthorhombic, monoclinic, or triclinic optical symmetries.

Applied to a F1-STF, the optical equivalent can be, in the most simple case, a single biaxial layer described by the HBLA. This biaxial layer comprises then an optical thickness $d$, corresponding to the actual thickness of the nanostructured thin film as well as external Euler angles ($\varphi$, $\theta$, $\psi$) and internal angles ($\alpha$, $\beta$, $\gamma$) determining the orientation of the columns and sample during a particular measurement and biaxial properties, respectively~\footnote{It is not a priori knowledge that the Euler angles, which diagonalize the HBLA tensor are equivalent to the intrinsic structural properties of the composite material such as slanting angle of the columns and orientation of the slanting plane relative to the external laboratory coordinate system. It has been confirmed by extensive investigations that Euler angles $\theta$ and $\varphi$ are identical to these properties, respectively.}. Furthermore, there are three independent, complex, and wavelength-dependent functions $\varrho(\omega)_{j}$, pertinent to major polarizability axes $j=\mathbf{a},\mathbf{b},\mathbf{c}$~\cite{Schmidt2009a,Schmidt2009b,Schmidt2013ellinanoscale}.

Explicitly, the dielectric tensor $\boldsymbol{{\varepsilon}}_{\textrm{t}}$ for a triclinic material takes the form
\begin{equation}\label{eq:monoclinicepstensor}
{\boldsymbol{\varepsilon}}_{\textrm{t}}=\mathbf{A}\mathbf{U} \left(
\begin{array}{ccc}
{\varrho(\omega)_{a} } & 0  & 0 \\
0  & {\varrho(\omega)_{b} } & 0 \\
0  & 0 & {\varrho(\omega)_{c} } \\
\end{array} \right) \mathbf{U}^{\rm t}\mathbf{A}^{\rm t},
\end{equation}
where $\mathbf{A}$ is the real-valued Euler angle rotation matrix and $\mathbf{U}$ is the projection matrix~\cite{Schmidt2013ellinanoscale}. Note that here the superscript ``t'' refers to the transpose of the respective matrix.

The HBLA does not allow to determine fractions of constituents within the composite material, nor the constituent bulk-like optical properties of the building blocks. However, the HBLA has several advantages over other effective medium approximations: (i) no initial assumptions such as optical parameters of the constituents or material fractions are necessary, (ii) it is valid for absorbing and non-absorbing materials, and (iii) it does not depend on the structure size. Note that the actual structure size is disregarded in this homogenization approach. This procedure is considered valid since the dimensions and especially the diameter of the nanostructures under investigation are much smaller than the probing wavelength. Care must be taken when properties at shorter wavelengths are evaluated, because diffraction and scattering phenomena may be present.

In general, it is presumed that the HBLA method together with the assumption of one effective optical thickness $d$ applied to match experimental data for a F1-STF delivers the best possible dielectric tensor $\boldsymbol{\varepsilon}$, i.e. $\varepsilon_{\textrm{eff},j}$ are considered the true effective major axes dielectric functions and therefore target functions for other effective medium approximations.

\section{Piecewise Homogeneous Biaxial Layer Approach}
If substrate rotation is involved during the growth process of STFs, a single biaxial layer accounting for the film is not sufficient anymore to describe the dielectric polarization response.
For the piecewise homogeneous biaxial layer approach two types of STFs are distinguished here: (i) F-STFs\footnote{For details on the nomenclature see Ref.~\cite{Schmidt2013ellinanoscale}} (except F1; fabricated with sequential substrate rotations) and (ii) H-STFs (fabricated with continuous substrate rotation). It is assumed that the STF is made of $m$ F1-STF slices, where within each slice (layer) the dielectric properties are homogeneous~\cite{LakhtakiaBook,Podraza2004}.

\begin{figure}[htbp]
\begin{center}
\includegraphics[width=0.8\linewidth, clip, trim=110 150 380 50]{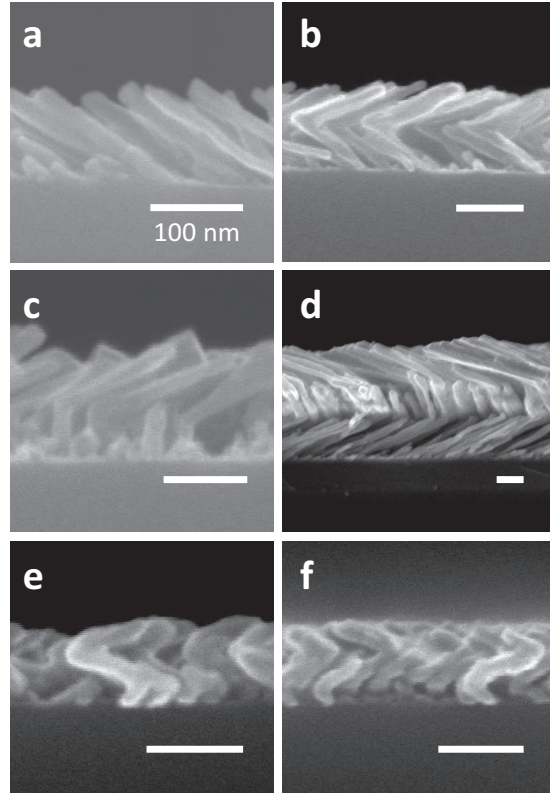}
\caption{High resolution cross-sectional SEM images of several achiral and chiral STFs: (a) Co F1-STF; (b) Ti 2F2-STF; (c) chiral Ti 2F4$^+$-STF; (d) chiral Ti 3F4$^+$-STF; (e,f) H$^+$ and H$^-$-STF, respectively. All scale bars are 100 nm.\label{SEMs}}
\end{center}
\end{figure}

\paragraph{F-STFs}
F-STFs (all but F1) are grown while the substrate is rotated step-wise (abruptly) after a certain growth time. If a sequential substrate rotation of $180^\circ$ is employed, for example, the resulting chevrons or zig-zags (F2-STFs), can be considered as stratified (or a cascade of) F1-STFs with opposite slanting directions in adjacent slabs. Consequently, the optical model equivalent for a chevron thin film with two legs (2F2-STF, Fig.~\ref{SEMs}b) may consist of two homogeneous anisotropic (biaxial) layers on top of a layer accounting for the substrate. The Euler angles for both layers ($\varphi_j$,$\theta_j$,$\psi_j$), which transform the Cartesian coordinate system $(x,y,z)$ into the sample coordinates $(\xi, \eta, \zeta)$, represent the orientation of each slanted column (building block) within the nanostructure.
In case the angle of the incoming particle flux $\theta_\textrm{i}$ was kept constant during deposition, a common dielectric tensor, with three major polarizabilities $\varrho_a$, $\varrho_b$, $\varrho_c$ pertinent to the intrinsic axes $\mathbf{a}$, $\mathbf{b}$, $\mathbf{c}$, internal angles $\alpha, \beta, \gamma$, and Euler angles $\varphi,\theta,\psi$ can be assigned to each biaxial layer. Deposition at constant $\theta_\textrm{i}$ results in equal packaging fractions in subsequent layers and therefore common effective major polarizabilities may be assumed. Furthermore, both layers have an individual thickness parameter $d_j$ such that the total thickness is equal to the overall film thickness ($d=d_1+d_2$). This approach is valid, in general, for arbitrarily achiral and chiral STFs which can be subdivided into stratified F1-STFs~\cite{Schmidt2013ellinanoscale}.

\paragraph{H-STFs}
If the substrate is continuously rotating around the normal during deposition, helical STFs (H-STFs, Fig.~\ref{SEMs}e,f) are growing since the sample rotation is equivalent to a constant angular change of the incoming vapor flux direction around the substrate normal and thus the self-shadowed regions change dynamically. H-STFs represent rotationally inhomogeneous anisotropic material with a twist along the sample normal and can be considered as ``frozen'' cholesteric liquid crystals~\cite{Schubert1996b,Schubert2001}. In order to model the electromagnetic plane wave response of H-STFs the thin film has to be virtually separated into $m$ homogeneous anisotropic layers with subsequently shifted Euler angle parameters $\varphi_1,\varphi_2,\ldots,\varphi_m$ with individual thickness parameters $\delta d = d/m$. These layers represent piecewise rotation with respect to each other by $\delta\varphi$ to resemble the twisted character. Physical quantities such as principal dielectric functions (as a function of photon energy and $z$), orientation, overall thickness, and handedness can be thereby retrieved from the ellipsometry model calculations. In contrast to F-STFs, for H-STF the Euler angle $\psi$ is found to be not equal to zero.

For a more detailed description and several examples of achiral and chiral STFs the interested reader is referred to Ref.~\cite{Schmidt2013ellinanoscale}, for example.

\section{Bruggeman Formalisms}
The Bruggeman formalism is a homogenization approach with absolute equality between the constituents in a mixture, and was originally developed for a medium comprising two randomly distributed spherical dielectric components~\cite{Bruggeman1935}. This isotropic Bruggeman formula has been extensively discussed and generalized to treat materials with multiple anisotropic constituents by introducing so-called depolarization factors, which are functions of the shape of the inclusions~\cite{Polder1946,Stroud1975,Granqvist1977,Mackay2012}. For ellipsoidal particles however, two different modifications of the Bruggeman formalism were suggested, which differ in the definition of these depolarization factors.

The generalization of the Bruggeman formalism with a definition of the depolarization factors introduced to optics by Polder and van Santen (Eq.~\ref{eq:biaxBrugge1b}) has been extensively used and applied to the analysis of experimentally acquired data of anisotropic thin films~\cite{Polder1946,Granqvist1977,Smith1989,Mbise1997,Beydaghyan2005,Nerbo2010,Hofmann2011,Wakefield2011,Schmidt2012a,Schmidt2013ellinanoscale}. This formalism will be called henceforth ``traditional anisotropic Bruggeman effective medium approximation'' (TAB-EMA). The implicit TAB-EMA formulae for the three effective major dielectric functions $\varepsilon_{\stext{{eff},\textit{j}}}^{\stext{T}}$ with $j=a,b,c$ for a mixture of $m$ constituents with fractions $f_n$ and constituents bulk-like dielectric functions $\varepsilon_n$ are
\begin{equation}
\label{eq:biaxBrugge1b}
\sum\limits_{n = 1}^m f_n\frac{\varepsilon_{n}-\varepsilon_{\stext{{eff},\textit{j}}}^{\stext{T}}}{\varepsilon_{\stext{{eff},\textit{j}}}^{\stext{T}} + L^{\stext{D}}_{\stext{\textit{j}}}(\varepsilon_{n} - \varepsilon_{\stext{{eff},\textit{j}}}^{\stext{T}})}=0,
\end{equation}
with the depolarization factors
\begin{equation}
\label{eq:Brugge1depol}
L^{\stext{\rm D}}_{j}=\frac{U_xU_yU_z}{2}\int\limits_0^\infty\frac{(s+U_j^2)^{-1}{\rm d}s}{\sqrt{(s+U_x^2)(s+U_y^2)(s+U_z^2)}}.
\end{equation}
The definition of $L^{\stext{\rm D}}_{j}$ is based on the potential of uniformly polarized ellipsoids and has been adapted from magnetostatic theory where these parameters are well-known under the name `demagnetizing factors'~\cite{Osborn1945}. It is important to note that the real-valued depolarization factors $L^{\stext{\rm D}}_{j}$ only depend on the real-valued shape parameters $U_j$ of the ellipsoid and that the two ratios ($U_x/U_z$) and ($U_y/U_z$) serve to define the shape exactly. It can be shown that the depolarization factors of an ellipsoid satisfy the relation
\begin{equation}
\label{eq:unity}
L^{\stext{\rm D}}_{x}+L^{\stext{\rm D}}_{y}+L^{\stext{\rm D}}_{z}=1.
\end{equation}
Furthermore, the sum of all $f_n$ has to equal unity.
Analytical solutions for Eq.~(\ref{eq:biaxBrugge1b}) still exist even with several constituents $m$ and the physically correct solution of the polynomial equation can be determined by an algorithm based on conformal mapping, for example~\cite{Jansson1994}.

The second existing Bruggeman formalism comprises depolarization factors that are based on Green functions and was first introduced by Stroud in 1975~\cite{Stroud1975}. Recently, Mackay and Lakhtakia published explicit equations for these depolarization factors for the case of anisotropic inclusions~\cite{Mackay2012}. The effective permittivity parameters $\varepsilon_{\stext{{eff},\textit{j}}}^{\stext{R}}$ are given implicitly by the three coupled equations
\begin{equation}
\label{eq:biaxBrugge2b}
\sum\limits_{n = 1}^m f_n\frac{\varepsilon_{n}-\varepsilon_{\stext{{eff},\textit{j}}}^{\stext{R}}}{1 + D^{\stext{D}}_{\stext{\textit{j}}}(\varepsilon_{n} - \varepsilon_{\stext{{eff},\textit{j}}}^{\stext{R}})}=0,
\end{equation}
with the depolarization factors specified by the double integrals
\begin{subequations}\label{eq:Brugge2depola}
\begin{align}%
D^{\stext{\rm D}}_{x}&=\frac{1}{4\pi}\int\limits_{0}^{2\pi}\int\limits_{0}^{\pi}\frac{\sin^3\vartheta\cos^2\phi}{U_x^2\rho}{\rm d}\vartheta{\rm d}\phi\label{first},\\
D^{\stext{\rm D}}_{y}&=\frac{1}{4\pi}\int\limits_0^{2\pi}\int\limits_0^{\pi}\frac{\sin^3\vartheta\sin^2\phi}{U_y^2\rho}{\rm d}\vartheta{\rm d}\phi\label{second},\\
D^{\stext{\rm D}}_{z}&=\frac{1}{4\pi}\int\limits_0^{2\pi}\int\limits_0^{\pi}\frac{\sin\vartheta\cos^2\phi}{U_z^2\rho}{\rm d}\vartheta{\rm d}\phi\label{third},
\end{align}
\end{subequations}
which involve the scalar parameter
\begin{equation}
\label{eq:Brugge2rho}
\rho=\frac{\sin^2\vartheta\cos^2\phi}{U_x^2}\varepsilon_{\stext{{eff},\textit{x}}}^{\stext{R}}+\frac{\sin^2\vartheta\sin^2\phi}{U_y^2}\varepsilon_{\stext{{eff},\textit{y}}}^{\stext{R}}+\frac{\cos\vartheta}{U_z^2}\varepsilon_{\stext{{eff},\textit{z}}}^{\stext{R}}.
\end{equation}
The depolarization factors $D^{\stext{\rm D}}_{j}$ are, in general (lossy medium), complex parameters and are a function of the shape parameters $U_j$ of the ellipsoid as well as the effective permittivities $\varepsilon_{\stext{{eff},\textit{j}}}^{\stext{R}}$ of the medium. This formalism will be called henceforth ``rigorous anisotropic Bruggeman effective medium approximation'' (RAB-EMA). Note that due to the coupled nature of the RAB-EMA formalism generally numerical methods are necessary to calculate the effective permittivities $\varepsilon_{\stext{{eff},\textit{j}}}^{\stext{R}}$. In contrast to the TAB-EMA, the RAB-EMA has only been discussed mathematically and no reports on the application to evaluate experimentally acquired data from anisotropic samples exist. Furthermore, it should be noted that both theories (i) presume structural equivalence for all $m$ constituents and (ii) are identical for the limiting case of isotropic spherical inclusions ($U_x=U_y=U_z$).

\section{Experimental}
The F1-STFs were deposited in a custom built ultra-high vacuum chamber by means of electron-beam evaporation onto Si(100) substrates with a native oxide. The chamber background pressure was in the $10^{-9}$~mbar range and the substrates were held at room-temperature during the fabrication through a water-cooled sample holder. Both samples discussed here have been deposited at a constant particle flux of approximately 4~${\rm \AA}/$s (measured at normal incidence) while the substrate normal was tilted away from the particle flux by $85^\circ$. Samples A and B consist of Co and NiFe slanted columns, respectively and have a nominal film thickness of 100~nm.

\begin{figure}[htbp]
\begin{center}
\includegraphics[width=0.8\linewidth, clip, trim=0 0 0 0]{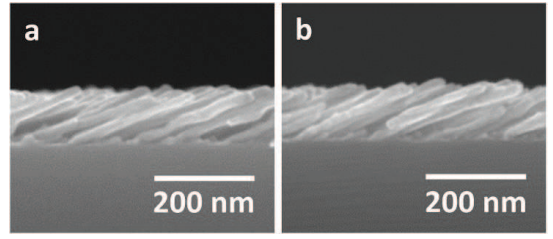}
\caption{High resolution cross-sectional SEM images of (a) an as deposited Co F1-STF and (b) the same sample after 50 cycles of alumina ALD at a substrate temperature of 80~$^\circ$C.\label{CoSEMs}}
\end{center}
\end{figure}

After each growth process the samples have been transferred to the ALD reactor (Fiji 200, CambridgeNanotech) and both samples have been coated with alumina by using trimethylaluminum and nanopure water as precursors. Sample A has been exposed to 50 precursors cycles with a substrate heater setting of 80~$^\circ$C and 45 cycles have been run at a substrate heater setting of 200~$^\circ$C for sample B. The purge times between the precursor pulses of 60~ms were 40~s and 15~s for samples A and B, respectively. Figure~\ref{CoSEMs} depicts a cross-sectional high-resolution SEM image of an identical sample to sample A before and after the conformal ALD layer was grown.

The ALD process for sample B has been monitored by \emph{in-situ} ellipsometry (M2000U, J.A. Woollam Co. Inc.) with an angle of incidence of 68$^\circ$. Since there is no sample azimuth rotation capability inside the ALD reactor the sample was placed on the chuck in such a manner that the long axis of the columns was rotated by approximately $45^\circ$ with respect to the plane of incidence. The acquisition time for one Mueller matrix spectrum ranging from 400 to 730~nm was approximately 20~s.

As soon as the samples had been taken out of the ALD reactor ex-situ Mueller matrix ellipsometry spectra within a spectral range from 400 to 1650~nm have been acquired at angles of incidence $\Phi_{\rm a}=45^\circ,55^\circ,65^\circ,75^\circ$ (M2000VI, J.A. Woollam Co. Inc.). Additionally, to allow for accurate evaluation of the sample anisotropy, at each angle $\Phi_{\rm a}$ spectra were measured over a full azimuthal rotation every six degrees.


\section{Results and Discussion}
\subsection{Optical model comparison}
The experimentally acquired Mueller matrix spectra of the Co F1-STF conformally coated with Al$_2$O$_3$ (sample A) have been analyzed, for reasons of comparison, with HBLA, TAB- and RAB-EMA optical models. In all three cases the assumption is made that the optical response of the F1-STF can be described by a single homogeneous biaxial layer~\cite{Schmidt2009a,Schmidt2009b,Schmidt2013ellinanoscale}. For both EMA models a composite material with three different constituents ($m=3$), one of them being void ($\varepsilon_1=1$), is assumed. Model parameters for the HBLA are film thickness $d$, Euler angles $\varphi$ and $\theta$, a monoclinic angle $\beta$, and the effective major axis dielectric functions $\varepsilon_{\textrm{eff},j}$. The TAB- and RAB-EMA model parameters comprise additionally two effective shape parameters ($U_x, U_y$) and Al$_2$O$_3$ ($f_{\rm Al2O3}$) and Co ($f_{\rm Co}$) fractions. Note that for both EMA models only the constituent bulk-like dielectric function $\varepsilon_{\rm Co}$ has been included in the best-match model analysis while for the transparent alumina a bulk-like dielectric function $\varepsilon_{\rm Al2O3}$ has been assumed, which was determined by ellipsometry prior to this investigation from a 18~nm thin solid film with the Cauchy dispersion model~\cite{Schmidt2012a}.

\begin{table}[tbp]
  \centering
  \caption{Summary of the best-match model parameters for the conformally alumina coated Co F1-STF determined with HBLA, TAB-EMA, and RAB-EMA model approaches.}
    \begin{tabular}{l|rrr}
    \hline\hline
    Parameter & HBLA & TAB-EMA & RAB-EMA  \\
    \hline
    $t$ (nm)            & 95.9(2)   & 92.34(7)   & 96.20(9)     \\
    $\theta$ ($^\circ$) & 58.38(3)  & 57.63(3)   & 58.33(2)     \\
    $\beta$ ($^\circ$)  & 89.1(2)   & 81.40(7)   & 89.73(8)     \\
    $f_{\rm Co}$ (\%)   & ---       & 22.9(2)    & 21.78(5)     \\
    $f_{\rm Al2O3}$ (\%)& ---       & 14.9(2)    & 19.40(7)     \\
    $U_x$               & ---       & 0.23(3)    & 0.118(3)     \\
    $U_y$               & ---       & 0.19(3)    & 0.078(2)     \\
    MSE                 & 10.66     & 15.13      & 11.42        \\
    \hline\hline
    \end{tabular}%
  \label{tab:parameters}%
\end{table}%

\begin{figure*}
\begin{center}
\floatbox[{\capbeside\thisfloatsetup{capbesideposition={right,center},capbesidewidth=5cm}}]{figure}[\FBwidth]
{\caption{Effective major axes optical constants, refractive indices $n_j$ and extinction coefficients $k_j$, along major polarizability axes $\mathbf{a}$, $\mathbf{b}$, $\mathbf{c}$ of sample A determined by HBLA (black solid lines), TAB-EMA (blue dashed lines) and RAB-EMA (red dotted lines) (top row) and corresponding constituent bulk-like optical constants, refractive indices and extinction coefficients, and also in comparison with data obtained from a 100~nm thin Co solid film (bottom row).}\label{fig:nk}}
{\includegraphics[width=12cm, clip, trim=0 30 0 60]{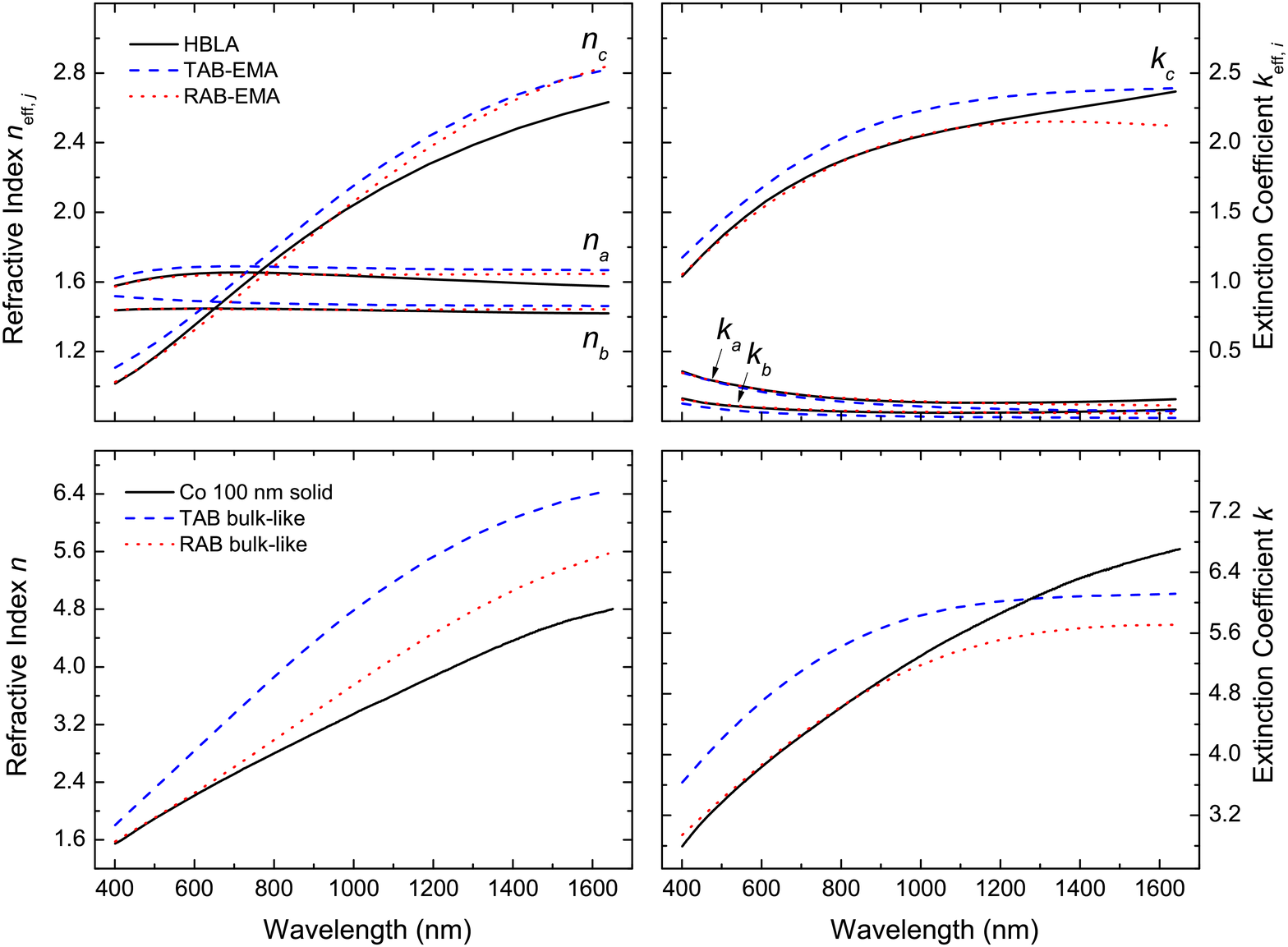}}
\end{center}
\end{figure*}

Best-match model structural parameters for all three optical models are listed in Table~\ref{tab:parameters}.
The HBLA model results are in very good agreement with SEM image analysis and the effective major axes optical constants show the expected birefringence and dichroism trends known from similar previous investigations~\cite{Schmidt2009c,Schmidt2012a,Schmidt2013ellinanoscale}. The EMA formalisms are in very good agreement with the HBLA in terms of film thickness $d$ and especially slanting angle $\theta$. Only the TAB-EMA suggests a significant monoclinic angle in contrast to HBLA and RAB-EMA. Furthermore, both EMA formalisms deliver fairly consistent results with respect to the Co fraction $f_{\rm Co}$ and calculated void fraction values between 75\% and 80\% are also in agreement with general trends of porosity values for samples deposited at such glancing angles~\cite{Beydaghyan2005}. The shape parameters ($U_x$, $U_y$) should ideally render the geometry of the core-shell columns, which can only be approximated as ellipsoids. The fact that $U_z=1\neq U_x\neq U_y$ constitutes biaxial film properties and the larger $U_z$ is with respect to $U_x$ and $U_y$ the more elongated is the rendered ellipsoidal particle. Here, both EMAs deliver reasonable estimates of the true column's aspect ratio of about 7. Note that since the slanted columns interact with the substrate and with each other and an ellipsoidal shape is only a rough approximation, $U_j$ should be considered as effective shape factors that may not necessarily be representative of the true geometry of the inclusions~\cite{Granqvist1977}.
The most desired parameter of the EMA analysis is the Al$_2$O$_3$ constituent film fraction and which differs by almost 5\% between the two anisotropic Bruggeman formalisms. With simple geometric considerations based on best-match model results and an average column diameter of 21~nm determined from top-view and cross-section SEM images the surface area can be obtained, which allows for the conversion of the alumina fraction parameter into a layer thickness~\cite{Schmidtdiss}. Hence, the TAB- and RAB-EMA values of 14.9\% and 19.4\% correspond to Al$_2$O$_3$ layer thicknesses of 2.7 and 3.7~nm, respectively. The layer thickness estimate based on the TAB-EMA results is in excellent agreement with a flat reference sample, which shows an alumina thickness on top of a thin solid Co film of 2.74~nm measured by standard ellipsometry as well as cross-sectional SEM image analysis, which reveals an average column diameter increase of 4.2~nm.

Figure~\ref{fig:nk} depicts effective major axes dielectric functions determined by HBLA ($\varepsilon_{\stext{{eff},\textit{j}}}=(n_{\stext{{eff},\textit{j}}}+ik_{\stext{{eff},\textit{j}}})^2$), TAB-, and RAB-EMA as well as the corresponding constituent bulk-like dielectric functions of Co ($\varepsilon_{\rm Co}=(n_j+ik_j)^2$) in comparison to the bulk dielectric function obtained from a 100~nm thin Co solid film.
The effective optical constants along the three major polarizability axes determined by the HBLA are considered as true values and therefore standard for the respective TAB and RAB results. While the TAB-EMA formalism overestimates the effective refractive indices and slightly deviates from the extinction coefficients within the investigated spectral region the RAB-EMA formalism shows a good agreement with the HBLA within the visible spectrum and only deviates in the long wavelength range.

The corresponding constituent bulk-like optical constants resulting from matching TAB-EMA and RAB-EMA to experimental Mueller matrix spectra are significantly different from each other and also from data obtained from the 100~nm thin solid reference film (Fig.~\ref{fig:nk}). Only within the visible wavelength range the RAB-EMA data show good resemblance with the respective bulk reference data. Differences between bulk material reference optical constants and constituent bulk-like optical constants of such F1-STFs determined with the two Bruggeman EMA formalisms discussed here may have several origins. First of all, the dielectric properties of ultrathin metal films may differ from their respective bulk properties due to surface and quantum confinement effects, which is very well possible here considering isolated columns with diameters of around 20~nm and less~\cite{Oates2004,Hövel2010,Alonso2010}. It has been recently shown that the thin conformal dielectric surface passivation layer affects the bulk-like optical properties of the Co core possibly due to the large surface to volume ratio of about 200~m$^{-1}$~\cite{Schmidt2012b}. Another consideration is that both Bruggeman formalisms are based on an idealized model of randomly distributed ellipsoidal particles and this description differs from the sample under investigation, which consists of columns with approximately elliptical cross-section attached to a substrate. Additionally, the optical model equivalent of a nanostructured thin film is a single anisotropic layer, which neglects non-idealities due to a ``surface roughness'' and a very thin nucleation layer.

In general, it can be said that simply by considering the MSE the HBLA always delivers the best match between model and experimental data due to the independent determination of the effective optical constants along major polarizability axes. This observation is in accordance with the initial assumption that the HBLA method will deliver the best possible dielectric tensor.

\begin{figure*}[htbp]
\begin{center}
\includegraphics[width=0.9\linewidth, clip, trim=0 30 0 0]{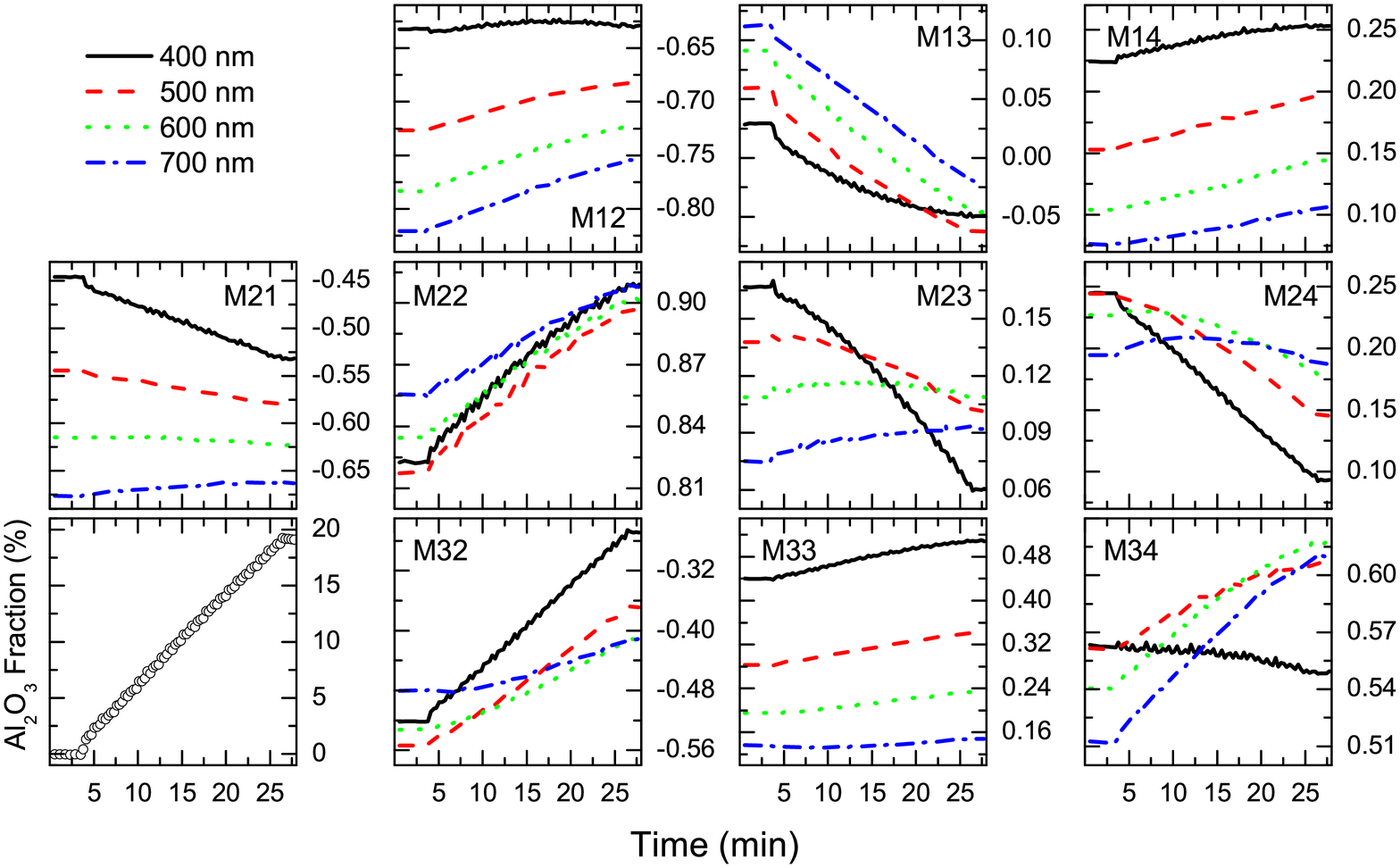}
\caption{Selected Mueller matrix elements with experimental data at four different wavelength plotted versus time reflect the changes in the optical response of the NiFe F1-STF during 45 cycles of alumina ALD. The first trimethylaluminum (TMA) precursor pulse occurred at $t=3$:45~min. The lower left graph shows the best-match model result for the Al$_2$O$_3$ film fraction.\label{insitu}}
\end{center}
\end{figure*}

\subsection{\emph{In-situ} growth monitoring}
Having the ability of determining individual constituent fractions within an optically biaxial material with voids the TAB-EMA, for example, can then be used for \emph{in-situ} growth monitoring. An almost ideal scenario for such investigations is ALD since the layer-by-layer growth mode takes place on a comparable time-scale as the Mueller matrix ellipsometry measurements. Furthermore, the conformal characteristic, i.e. the excellent growth homogeneity on three-dimensional objects, continues to allow for the assumption of a single homogeneous biaxial model.
Figure~\ref{insitu} depicts the measured upper $3\times4$ Mueller matrix normalized to $M_{11}$ (except $M_{31}$) with respect to time for a 45 cycle alumina ALD process. Each Mueller matrix element shows four experimentally determined graphs, which depict different wavelengths (400, 500, 600, and 700~nm). The first TMA precursor pulse occurred at $t=3$:45~min, and which can be seen in the initial step-like change in certain graphs. The fact that the graphs look noisy stems from the cyclic nature of the growth process and the measurement time being longer than a half-cycle. Most of the depicted graphs show a linear behavior with progressing alumina layer growth, however in contrast to observations made when filling pores with toluene, not all block-off diagonal elements ($M_{13}$,$M_{14}$,$M_{32}$,$M_{34}$) linearly decrease in amplitude with decreasing void fraction~\cite{May2010}.

The lower left graph shows the best-match model alumina fraction parameter $f_{\rm Al2O3}$ plotted versus time and the excellent linear behavior is characteristic for the self-limiting layer-by-layer ALD growth. The final volume fraction of $f_{\rm Al2O3}=19.1\%$ translates into a conformal layer thickness of 2.4~nm  with the assumption of an average column diameter of 16~nm. This value is in very good agreement with the alumina layer thickness of 2.21~nm determined by best-match model analysis of the ex-situ Mueller matrix spectra acquired at multiple angles of incidence and multiple in-plane orientations as well as a flat reference sample, which shows an alumina thickness on top of a thin solid NiFe film of 2.63~nm. Note that the average column diameter is determined by SEM image analysis and hence the conformal layer thickness values can have error bars of up to 20\%.



\section{Conclusions}
Optical model strategies to analyze generalized Mueller matrix ellipsometry spectra of ultrathin sculptured thin films (STFs) with simple and complex geometries based on the homogeneous biaxial layer approach (HBLA) have been presented. Additionally, two different anisotropic effective medium approximations (EMA) formalisms originating from the Bruggeman equation for spherical inclusions have been discussed. The determination of structural and effective optical properties with both EMA formalisms and the HBLA approach has been illustrated on the example of cobalt F1-STFs conformally coated with alumina by atomic layer deposition. Structural parameters, film thickness and column slanting angle, obtained by best-match regression analysis of experimental data with the EMA optical models are in very good agreement with values determined by the HBLA. Most importantly, the additional information gained by applying the EMA formalisms, the film constituent fractions are in very good agreement with data obtained from isotropic reference samples and estimates from SEM image analysis. It was discussed, however, that if an accurate determination of the effective major dielectric functions is the desired result, the HBLA model needs to be used. Furthermore, constituent bulk-like dielectric functions of the metal core resulting from the EMA model analysis may differ from the respective bulk material dielectric function due to surface and confinement effects, for example.

Finally, the TAB-EMA model approach has been applied to \emph{in-situ} growth monitoring of conformal alumina coatings on permalloy F1-STFs. The analysis of Mueller matrix spectra revealed the expected linear growth characteristic of atomic layer deposition and final values are in very good agreement with ex-situ measurement analysis and a thin solid reference sample.

\section*{Acknowledgments}
D.S. gratefully acknowledges support from and fruitful discussions with Mathias Schubert, Eva Schubert, Tino Hofmann (all University of Nebraska-Lincoln), and Craig M. Herzinger (J.A. Woollam Co. Inc.) and financial support from the National Science Foundation in RII (EPS-1004094), CAREER (ECCS-0846329), and MRSEC (DMR-0820521), the University of Nebraska-Lincoln, and the J.A. Woollam Foundation.












\end{document}